\documentstyle[12pt,cite,epsf]{article}
\def\Thebibliography#1{\section*{References}
 \list
 {[\arabic{enumi}]}{
\settowidth\labelwidth{[#1]}\leftmargin\labelwidth
 \advance\leftmargin\labelsep
 \usecounter{enumi}}
 \def\newblock{\hskip .11em plus .33em minus .07em}
 \sloppy\clubpenalty4000\widowpenalty4000
 \sfcode`\.=1000\relax}

\def\s0{\setcounter{equation}{0}\setcounter{table}{0}\setcounter{figure}{0}}
\let\huge=\Large
\let\Large=\large
\let\large=\normalsize
\newcommand{\eqn}[1]{(\ref{#1})}
\newcommand{\be}{\vskip0.2truecm\begin{equation}}
\newcommand{\ee}{\end{equation}\vskip0.2truecm\noindent}
\newcommand{\bea}{\vskip0.2truecm\begin{eqnarray}}
\newcommand{\ea}{\end{eqnarray}\vskip0.2truecm\noindent}
\newcommand{\bp}{\mbox{\boldmath $p$}}
\newcommand{\bx}{\mbox{\boldmath $x$}}
\newcommand{\bC}{\mbox{\boldmath $C$}}
\newcommand{\bL}{\mbox{\boldmath $L$}}
\newcommand{\vsec}[1]{\vskip1.0truecm\section{#1}}

\newcommand{\cM}{{\cal M}}
\newcommand{\bel}[1]{\be\label{#1}}

\def\apj#1#2#3{          {\it Astrophys. J. }{\bf #1} (19#2), #3}
\def\nat#1#2#3{          {\it Nature }{\bf #1} (19#2), #3}
\def\aa#1#2#3{          {\it Astron. \& Astroph. }{\bf #1} (19#2), #3}
\def\apjs#1#2#3{         {\it Astrophys. J. Suppl. }{\bf #1} (19#2), #3}
\def\pr#1#2#3{           {\it Phys. Rev. }{\bf #1} (19#2), #3}
\def\pl#1#2#3{           {\it Phys. Lett. }{\bf #1} (19#2), #3}
\def\np#1#2#3{           {\it Nucl. Phys. }{\bf #1} (19#2), #3}
\def\prl#1#2#3{          {\it Phys. Rev. Lett. }{\bf #1} (19#2), #3}
\begin{document}
\pagestyle{plain}
\baselineskip=18pt
\parskip 0.2truecm
\vbadness=10000
\hbadness=10000
\tolerance=10000
%

% maggiore circa e minore circa%%%%%%%%%%%%%%%
\def\ltsima{$\; \buildrel < \over \sim \;$}
\def\simlt{\lower.5ex\hbox{\ltsima}}
\def\gtsima{$\; \buildrel > \over \sim \;$}
\def\simgt{\lower.5ex\hbox{\gtsima}}
\begin{flushright}
{\ submitted to {\it Astron. \& Asproph.}}
\end{flushright}

\centerline{\huge{\bf Phase--space Distribution of Volatile Dark Matter.}}
\vskip 1truecm
\centerline{S.~Ghizzardi,  S.A.~Bonometto\footnote{
e--mail:bonometto@astmiu.mi.astro.it}}
\vskip 0.5truecm
\centerline{Dipartimento di Fisica dell'Universit\'a di Milano,}
\centerline{ Via Celoria 16, I-20133 Milano, Italy}
\vskip 0.5truecm
\centerline{I.N.F.N. -- Sezione di Milano}

\begin{abstract}
We discuss the phase--space distribution of $\mu$ neutrinos if $\tau$
neutrinos are unstable and decay into $\nu_\mu + scalar$.
If this scalar is a familon or a Majoron, in the generic case the
$\nu_\mu$ background is NOT the straightforward overlap of neutrinos
of thermal and decay origins. A delay in $\nu_\tau$ decay,
due to the Pauli exclusion principle, can modify it in a significant way.
We provide the equations to calculate the $\nu_\mu$ distribution
and show that, in some cases, there exists a good approximate
solution to them. However, even when such solution is not admitted,
the equations can be numerically solved following a precise pattern.
We give such a solution for a number of typical cases.
If $\nu_\mu$ has a mass $\sim 2$ eV and the see--saw argument holds,
$\nu_\tau$ must be unstable and the decay into $\nu_\mu + scalar$
is a reasonable possibility. The picture leads to a delayed equivalence
redshift, which could allow to reconcile COBE data with a bias parameter
$b\ge 1$.
\end{abstract}

\vsec{Introduction}
Cosmological models based on cold dark matter (CDM) allow to approach
observational data on large scale structure (LSS) over a wide range of scales.
The residual discrepancy between CDM predictions and LSS data can be
covered by replacing CDM with a mix of various components, among whom
CDM is however the most relevant one. This is in line with the
ideas which led to mixed dark matter (MDM) models in the middle
eighties \cite{bv:85,vb:85,aos:85,hol:89},
i.e. that the Universe contains a number of particle
species in different proportions. Most of such species decoupled early
and most decoupled components were already non--relativistic when the galaxy
mass-scale entered the horizon. Some other component(s) became non
relativistic only later on. The former components are CDM and no LSS
observation discriminates among them. The latter one(s) will be called
{\sl volatile}; among them, {\sl hot} components are characterized by a phase
space distribution of thermal origin. Quite at variance from CDM, LSS
observations can discriminate among different mixtures of volatile components.
As an example, CHDM models with $\Omega_c= 0.6$, $\Omega_h= 0.3$,
$\Omega_b= 0.1$, $g_\nu=2$ ($\Omega$ are the usual density parameters;
the indeces $c,h,b$ refer to CDM, hot dark matter, and baryons, respectively;
$g_\nu$ is the number of spin states of the hot component) and C$2\nu$DM
models with $\Omega_c= 0.75$, $\Omega_h= 0.2$,
$\Omega_b= 0.05$, $g_\nu=4$ lead to substantially different predictions on
a number of statistical estimations, as the expected number of damped
$Ly_\alpha$ clouds and the void probability function (VPF)
\cite{seb}. Another example
is given by Pierpaoli \& Bonometto \cite{pierpa},
who considered different MDM models
with a volatile component originating from heavier particle decays, whose
transfer functions are substancially different both from CHDM and C$2\nu$DM,
but still agree with observational constraints for suitable parameters
choices.

Heavy particle decays can give origin to volatile components, whose
distribution was already studied \cite{bgm:94}.
A peculiar case however arises when
some decay products have the same nature of a pre--existing background
of thermal origin. The shape of the resulting distributions
can be significantly different from an overlap of thermal and
decay components and,
for a number of
cases, will be discussed in this paper.

A physical context in wich such problem is relevant arises if $\mu$ neutrino
has a mass $\sim 2$ eV \cite{phkc} and 3 lepton families
exist with neutrino masses related by see--saw formulae
\cite{gr:94,yana:79,gm:79,jrs:86,glas:91}.
The $\tau$ neutrino must then be unstable, not to over--close the Universe (the
same argument holds for any value $m_{\nu_\mu} \simgt 0.3$ eV). Its decay into
3 lighter $\nu$'s is however unlikely, as -- even with weak flavor  mixing --
it is forbidden at tree--level, unless {\sl ad hoc} models are built.
The decay modes  $\nu_\tau \to  \nu_{\mu,e} + e^+ + e^-$, instead, require
that $m_{\nu_\tau} \simgt 1$ MeV and would however seriously modify big--bang
nucleosynthesis (BBNS). A decay mode
\bel{eq:decay}
{\nu_\tau}\to \nu_\mu + \phi
\ee
is therefore the most reasonable possibility: $\phi$ is a scalar particle,
arising from the spontaneous breaking of some global symmetry.
Such scalars have been widely considered in the literature. Current examples
are Majorons and Familons
\cite{glas:91,GNY:83,wil:82,tao:94,wgs:94,cmp,schv:82,comp}.

According to \cite{wgs:94},
physical parameters are likely to yield a $\nu_\tau$
life ($\tau$) greater than BBNS age.
Accordingly, $\nu_\tau$ could become non--relativistic only when
$T \simlt 10$ KeV and decay when the neutrino temperature is $T_{dy} (\ll
m_{\nu_\tau})$.
This does not violate BBNS constraints. Decay yields are $\nu_\mu$
(possibly, but unlikely, also $\nu_e$) and scalar $\phi$'s.
When $\nu_\tau$ decay is essentially over, the Universe contains two
``overlapped'' $\nu_\mu$ backgrounds; $\nu_\mu$ of thermal origin derelativize
when $3T_\nu \simeq m_{\nu_\mu}$; $\nu_\mu$ originated from $\nu_\tau$
decay derelativize later, when $T_\nu \simeq m_{\nu_\mu}\left( T_{dy} /
m_{\nu_\tau}\right)$. Furthermore the Universe will contain electron neutrinos,
of negligible mass, and massless $\phi$'s, which have the same
behaviour of massless $\nu$'s in the shaping of LSS.
%%%%%%%%%%%%%%%%%FIGURE rad density {file= schema.ps}%%%%%%%%%%%%%%%%%
\begin{figure}
\vfill
\centerline{\mbox{\epsfysize=9.0truecm\epsffile{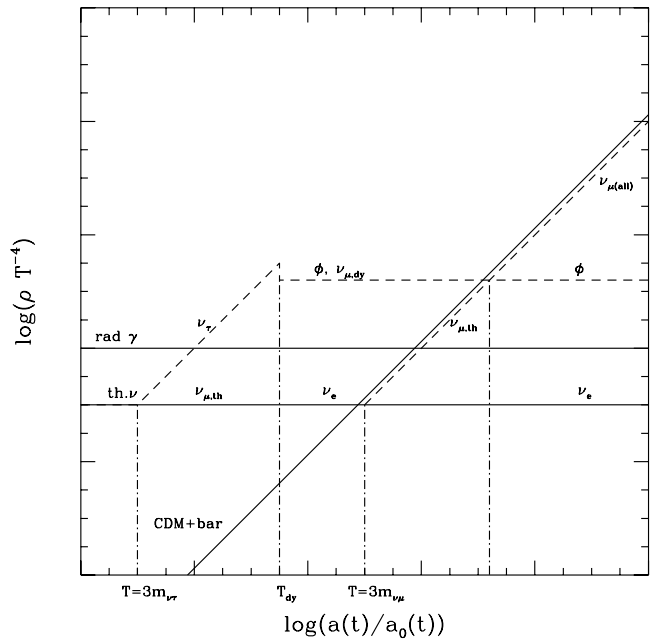}}}
\caption{\it The energy densities of the different components as
functions of the scale factor. In the actual world all ``corners'' should be
smoothed but this qualitative plot outlines: (i) The possible contribution of
scalars to the present massless background. (ii) The intermadiate behaviour
of the two $\nu_\mu$ backgrounds whose present number density -- and,
henceforth, energy -- is equal. Units on both axes are arbitrary.}
\label{fig:dens}
\end{figure}
%%%%%%%%%%%%%%%%%%%%%%%%%%%%%%%%%%%%%%%%%%%%%%%%%

In fig. \ref{fig:dens} we
plot the typical time dependence of the energy densities of radiation,
neutrinos, $\phi$'s and non--relativistic components. This figure stresses the
fact that $T_{\nu_\tau,dy}/T_{\nu_\mu,der} \simeq m_{\nu_\tau}/m_{\nu_\mu}$
($T_{\nu_\tau,dy}$: temperature at the time $\tau$ of $\nu_\tau$ decay;
$T_{\nu_\mu,der}$: temperature at $\nu_\mu$ derelativization).

Such scenario seems the most likely one, if $m_{\nu_\mu} \sim 2$ eV and
masses are related by the see--saw relation.
We shall now discuss the shape of the overall $\nu_\mu$
distribution in this case.
This is a typically non-thermal distribution and in the generic cases it is not
given by the straightforward overlap of decay and thermal distributions.
The final energy of decay $\nu_\mu$'s is enhanced by this effect.

\vsec{Coupled equations for decaying and daughter particles}
The phase space distribution $f(\bx, \bp, t)$ of light $\nu$'s (both of decay
and thermal origin) is obtainable from the Boltzmann equations
$\hat {\bL} [f]=\hat {\bC} [f]$,
where
\bel{eq:liouop}
\hat {\bL}= p^\alpha {\partial \over \partial x^\alpha}- \Gamma^\alpha_{\beta
\gamma}p^\beta p^\gamma {\partial \over \partial p^\alpha}
\ee
(greek indeces run from 0 to 3; $\Gamma$ are Christoffel symbols)
and $\hat {\bC}$ is the decay operator.
Assuming homogeneity and isotropy, in a Robertson--Walker model with scale
factor $a(t)$, $f$ depends only on $p^0=E$ and on the time $t$. It is also
important to separate the time dependence with occurs because of the decrease
of the temperature $T$ when $a$ encreases from an actual evolution
due to particle events. In general,
\bel{eq:lomis}
\hat {\bL} \left[f(E,T)\right] = E {\partial f \over \partial t }-
{\dot a \over a}E^2 {\partial f \over \partial E }
\ee
and, if $f_{\nu_\tau}$ is the phase--space distribution
of the mother particle ($\nu_\tau$), for the decay mode \eqn{eq:decay},
\bel{eq:opcollnu}
\hat {\bC} [f]  =  8 \pi^4 \int {d\Pi_{\nu_\tau} d\Pi_\phi
\delta^{(4)}\left( p_{\nu_\tau} - p_{\nu_\mu} -p_\phi\right)}
\left[\left| \cM_{fi}\right|^2_{\nu_\tau \to \nu_\mu + \phi}
f_{\nu_\tau} (1 -f)\right],
\ee
with
$d\Pi_i = g_i {d {\bp}_i \over (2 \pi)^3 2 E_i}$
($i=\nu_\tau, \phi$; $g_i$ is the number of independent spin--states of the
corresponding particle).
Assuming $\tau$ neutrinos at rest, taking $x=p/T$,
and splitting $f(x,t)$ into the sum $f_{dy}(x,t)+fer(x)$ with
$fer(x)=2/(e^x+1)$, eqs. \eqn{eq:lomis} and \eqn{eq:opcollnu} yield
\bel{eq:clsi}
{d f_{dy}\over dt}(x,t)  =
{2 \pi^2 \over \tau x^2}
{n_{\nu_\tau}(t) \over T^3}
\delta \left( x-{m_{\nu_\tau} \over 2T}\right)
\left(1 -fer(x)-f_{dy}(x,t)\right) ~.
\ee
Here $n_{\nu_\tau}$ is the number density of
$\nu_\tau$'s at the time $t$ and
$\tau=16 \pi {m_{\nu_\tau}/ |{\cM}_{fi}|^2}$
is the $\nu_\tau$ mean--life, which can be obtained from the decay matrix
${\cM}_{fi}$ evaluated for $E_{\nu_\tau}=m_{\nu_\tau}$ and $p_{\nu_\mu}=
p_\phi = m_{\nu_\tau}/2$. In turn, $N_{\nu_\tau}= n_{\nu_\tau} a^3$ depends
on the decay rate, and is still related to $f_{dy}$ according to
\bel{eq:ncomf}
{dN_{\nu_\tau} \over dt}= -{dN_{dy}\over dt}=-{a^3 T^3 \over (2\pi)^3}
{d \over dt}
\int{ d^3 x f_{dy}(x,t)}
\ee
Taking eq. \eqn{eq:clsi} into account, eq. \eqn{eq:ncomf} becomes
\bel{eq:ncom}
{dN_{\nu_\tau}\over dt} =
-{1 \over \tau} N_{\nu_\tau}(t)
\left[ 1-fer\left({m_{\nu_\tau} \over 2T}
\right)
-f_{dy}\left({m_{\nu_\tau} \over 2T},t \right)\right]
\ee
If $fer\left({m_{\nu_\tau} \over 2T}\right)$ and $f_{dy}\left({m_{\nu_\tau}
\over 2T},t \right)$ are negligible, eq. \eqn{eq:ncom} gives the usual
exponential decay and the known $\nu_\mu$ spectrum.
Here we are interested in the opposite case, when such decay--inhibiting terms
cannot be disregarded. Neutrinos decoupled at the time $t_{dg}$ when the
temperature was $T_{dg} \sim 0.9$ MeV.
At that time
\bel{eq:coin}
{N_{\nu_\tau}(t_{dg})\over a^3 T^3}=
{N_{\nu_\tau}(t_{dg})\over a_{dg}^3 T_{dg}^3}=
{n_{\nu_\tau}(t_{dg})\over T_{dg}^3}={3 \zeta(3) \over 2 \pi^2}
\ee
and this equation provides us the initial conditions which allow to replace
eq. \eqn{eq:ncom} into eq. \eqn{eq:clsi}, so obtaining the equation:
\bel{eq:eqffinw}
{d f_{dy}\over dt}(x,t)  =
 6 \zeta(3){ {\tilde t}_{dg} \over \tau x}
\delta \left( t-{\tilde t}_{dg} x^2 \right)
\left(1-fer(x)-f_{dy}(x,t)\right)\times
$$$$
\times\exp { \left\{ -{1 \over \tau}
\int_{t_{dg}}^t {ds \left[1- fer\left(\sqrt{s/ {\tilde t}_{dg}}\right)
-f_{dy}\left(x_{dg}\left(s/ t_{dg}\right)^{1 \over 2},s\right)\right]}
\right\}}.
\ee
(here $x_{dg} = m_{\nu_\tau}/2 T_{dg}$ and
${\tilde t}_{dg}=t_{dg}/x_{dg}^2$), whose only unknown is $f_{dy}$.
Eq. \eqn{eq:eqffinw} is to be solved for all $x$, but is an ordinary
differential equation in respect to time. The $\delta$ function at its r.h.s.
allows us to write its formal integral as follows:
\bel{eq:solfin}
f_{dy}(x,t) =
6 \zeta(3) {{\tilde t}_{dg} \over \tau x}
\left\{1-fer(x)-
f_{dy}\left(x, {\tilde t}_{dg} x^2\right)\right\}
{}~\Theta\left(x-x_{dg}\right)
\Theta\left( t /{\tilde t}_{dg} -x^2\right)\times
$$$$
\times \exp{ \left\{ -{2 {\tilde t}_{dg} \over \tau }
\int_{x_{dg}}^{x}
{du ~u \left[ 1- fer(u)-f_{dy}\left(u,{\tilde t}_{dg} u^2\right)
\right]}\right\}}
\ee
(the $\Theta$'s are Heaviside step distributions).
In general, eq. \eqn{eq:ncom} and \eqn{eq:solfin} can be solved only
numerically. However, the non--differential form of \eqn{eq:solfin}
allows an easy numerical approach and,
in some cases, also an analytical solution can be found. Eq.
\eqn{eq:solfin} is one of the results of this work.

\begin{figure}
\vfill
\centerline{\mbox{\epsfysize=8.0truecm\epsffile{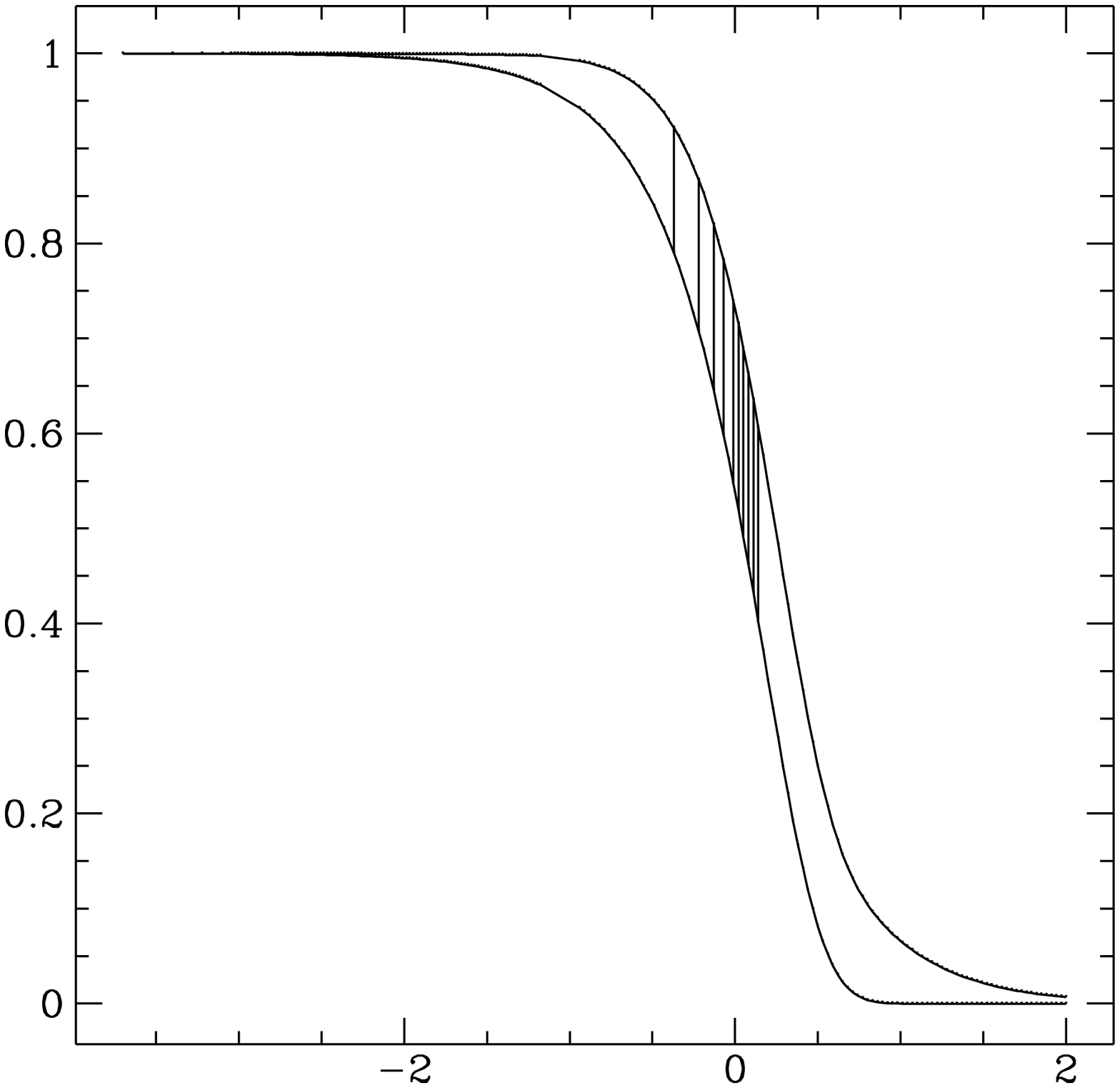}}}
\centerline{\mbox{\epsfysize=8.0truecm\epsffile{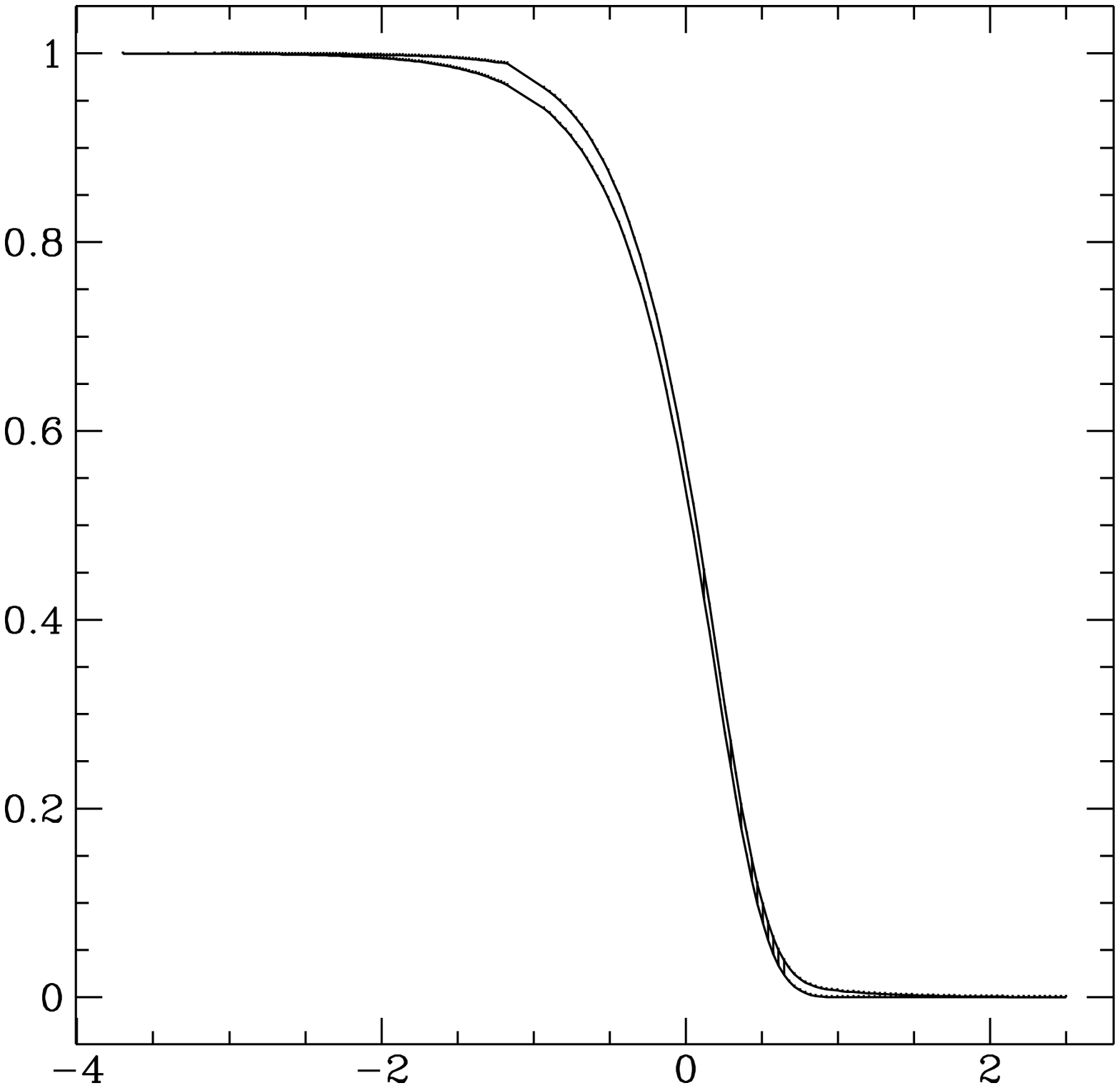}}}
\caption{\it Momentum distribution of $\nu_\mu$'s obtained
for $\tau=10^7$ sec and $\tau=10^8$ sec respectively. They are obtained
from eq. \protect\eqn{eq:solfin} and through the procedure indicated in
eqs. \protect\eqn{eq:sfdis} and \protect\eqn{eq:int}. In practice, no
approximation is made here and statistical effects have
been considered. In both plots the lower curve represents the preexisting
fermi distribution and the upper curve is the total distribution. In the
upper plot vertical lines indicate the shape of the spectrum at
intermediate times taken at constant time intervals.}
\label{fig:numsol}
\end{figure}

\vsec{Analytical and numerical solutions}
In order to solve eq. \eqn{eq:solfin}, let us consider a set of discrete times
$t_n =  t_{dg}+n~\Delta t$ and the corresponding values
$x_n = x_{dg} \left(t_n \over  t_{dg}\right)^{1\over 2}$. For $x_n$ in the
range set by the Heaviside distributions, eq. \eqn{eq:solfin} yields
\bel{eq:sfdis}
f(x_n,t_n)  =  6 \zeta(3) {{\tilde t}_{dg} \over \tau x}
\left(1-fer(x)-
f_{dy}\left(x_n, t_n\right)\right)\times
$$$$
\times\exp\left\{-{\Delta t \over \tau}
\sum_{k=1}^n
\left[1- fer(x_k) -f_{dy}
\left(x_k,t_k\right)\right]\right\}
\ee
For $n=1$, eq. \eqn{eq:sfdis} can be solved by intersecting the curves
$y=f$ and
\bel{eq:int}
y = 6 \zeta(3) {{\tilde t}_{dg} \over  \tau x_1 }
[ 1 - fer(x_1) - f ]
\exp \left\{ -{\Delta t \over \tau}
[1 - fer(x_1) - f] \right\}
\ee
(their intersection is unique).
Once $f(x_1,t_1)$ is obtained, we work out $f(x_2,t_2)$ in a similar fashion,
and, eventually, all $f(x_n,t_n)$.
By reducing the width of $\Delta t$, this approach allows any approximation and
a straightforward numerical algorithm built on this basis
is used to obtain figs. \ref{fig:numsol}.
Besides of the solutions obtained for two different parameter choices,
these figures also show the distribution obtained from the overlap of a
non-inhibited decay component with the thermal one.
This shows the significance of the inhibition effects.

\begin{figure}
\vfill
\centerline{\mbox{\epsfysize=8.0truecm\epsffile{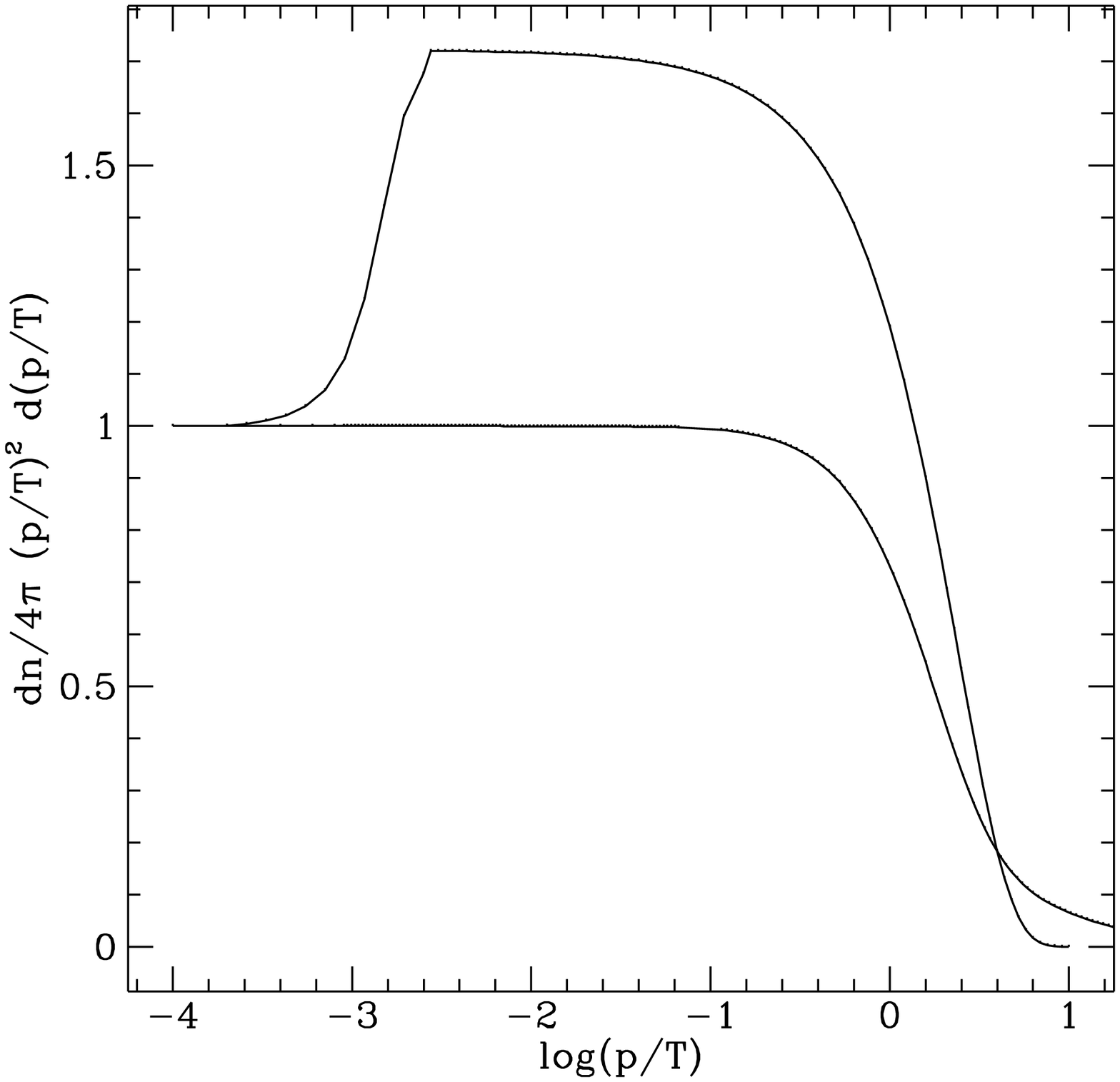}}}
\centerline{\mbox{\epsfysize=8.0truecm\epsffile{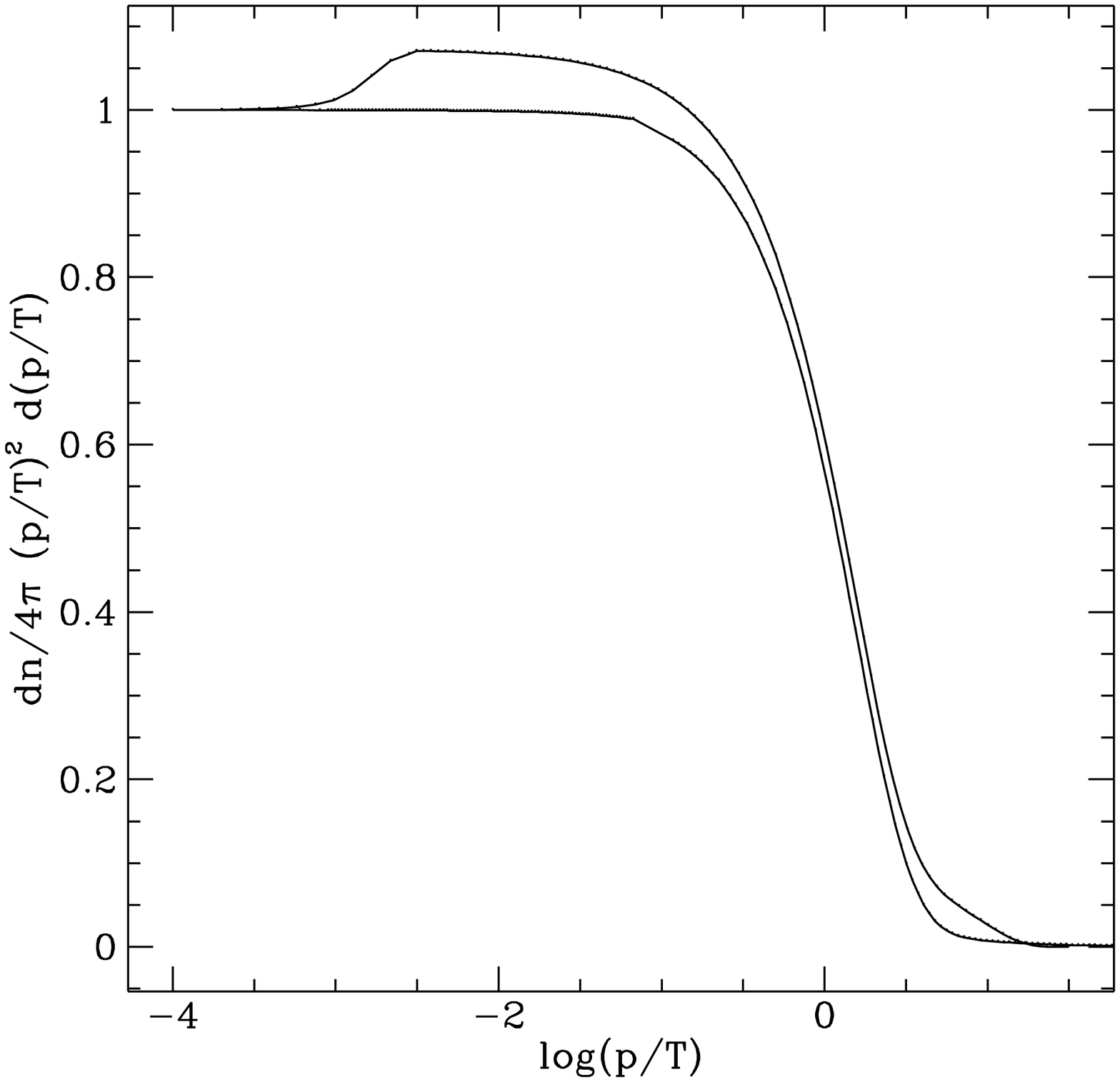}}}
\caption{\it
Total momentum distribution of $\nu_\mu$'s obtained
for $\tau=10^7$ sec and $\tau=10^8$ sec respectively.
They are obtained
from eq. \protect\eqn{eq:solfin} (upper plot) and from eq.
\protect\eqn{eq:solser} (lower plot). We compare the real solution with
the solution  obtained neglecting $f_{dy}$ in respect with $1-fer$.
For high $\tau$ the differences tend to disappear.}
\label{fig:diff}
\end{figure}

Analytical solutions can be found if $f_{dy}$ can be neglected in respect to
$1-fer$ and the lifetime $\tau$ is large enough to yield
$\tau x^2_{dg} / t_{dg} \gg 1$.
In such case eq. \eqn{eq:sfdis} is no longer a formal integral,
as $f_{dy}$ appears at the l.h.s. only.
Furthermore, taking into account that, for $y>0$,
$ ~\mbox{\rm tangh} (y/2) = 1 + 2 \sum_{n=1}^{+\infty}{(-1)^n e^{-ny}} $
and, henceforth,
\bel{eq:serp}
2\int_{x_{dg}}^x{dy \, y{{ e^y -1} \over {e^y +1}}}=
{x^2-x_{dg}^2}
 -\sum_{n=1}^{+\infty} \left[Z_n(x)-Z_n(x_{dg})\right]
\ee
with $Z_n(y)= 4(-1)^n \left( {y \over n} + {1 \over n^2}\right)e^{-ny}$, we
have that
\bel{eq:solser}
f_{dy}(x,t) =
6 \zeta(3) {t_{dg} \over \tau} {1 \over x x_{dg}^2}
[ 1 - fer(x) ]
\exp\left\{- {t_{dg} \over x_{dg}^2 \tau}
\left[(x^2-x_{dg}^2)
+\sum_{n=1}^{+\infty} [ Z_n(x) - Z_n(x_{dg}) ]
\right] \right\}
\ee
In the exponent the sum can be limited to the first $N$ terms, provided that
$N x_{dg} \gg 1$.
Such analytical integral is used to obtain fig. \ref{fig:diff} in which
we report both
solutions for two choices of parameters: as it is to be expected, taking
greater
values of $\tau$, the differences become
neglegible.
%%%%%%%%%%%%%%%%%FIGURE numer. sol. {files= numsol*.ps}%%%%%%%%%%%%%%%%%
%%%%%%%%%%%%%%%%%%%%%%%%%%%%%%%%%%%%%%%%%%%%%%%%%
%%%%%%%%%%%%%%%%%FIGURE two sol. {files=diff*.ps}%%%%%%%%%%%%%%%%%
%%%%%%%%%%%%%%%%%%%%%%%%%%%%%%%%%%%%%%%%%%%%%%%%%

\vsec{Discussion}
The expression \eqn{eq:solser} yields back the expression obtainable neglecting
the preexisting Fermi background, when $|\sum_{n=1}^{ \sim INT(10/x_{dg})} Z_n
(x)| \ll x^2$ for any $x > x_{dg}$ (here $INT(y)$ indicates the integer nearest
to $y$).
%%%%%%%%%%%%%%%%%FIGURE two sol. {files=inters.ps}%%%%%%%%%%%%%%%%%
\begin{figure}
\vfill
\centerline{\mbox{\epsfysize=4.0truecm\epsffile{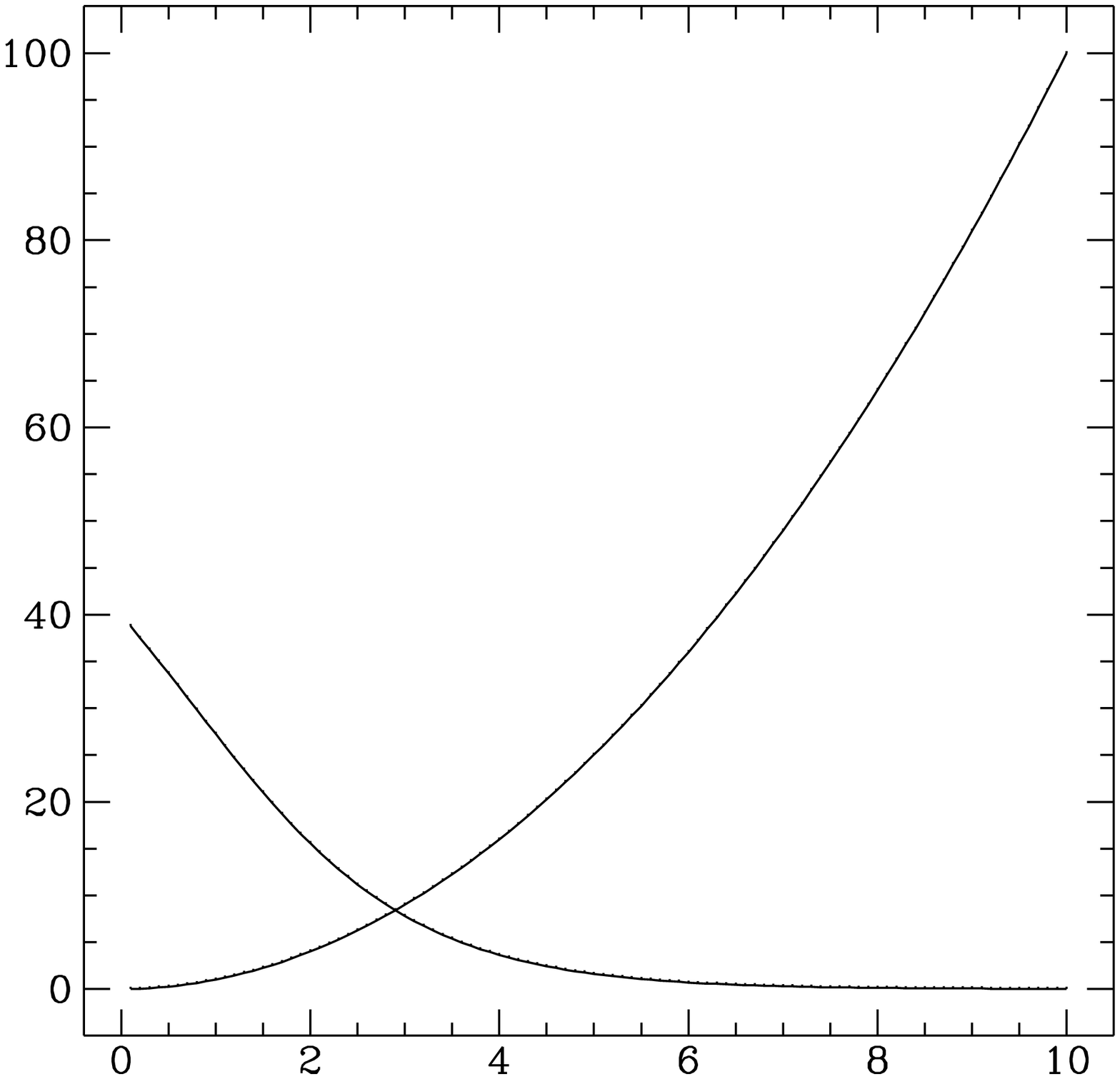}}}
\caption{\it Intersection between $10 \times |\sum_{n=1}^{ 3000 }
Z_n (x)| $ and $x^2$.}
\label{fig:inters}
\end{figure}
%%%%%%%%%%%%%%%%%%%%%%%%%%%%%%%%%%%%%%%%%%%%%%%%%
In fig. \ref{fig:inters} we plot the behaviour of
$10 \times |\sum_{n=1}^{ 3000 }
Z_n (x)| $ and $x^2$ (the upper limit is adequate if $m_{\nu_\tau} \simgt 5\,
$keV and therefore $x_{dg} \simgt 3 \times 10^{-3}$). The two curves intersect
for $x \simeq 3$. At greater $x$, the distance between the two curves widens
and the unequality holds for any $x > x_{dg}$ once it is true at $x_{dg}$. This
implies that the effects of the preexisting background cannot be neglected if
the $m_{\nu_\tau} < 6\, $MeV. According to the see--saw relation, if
$m_{\nu_\mu} \sim 2\, $eV, the neutrino distribution of thermal origin shall be
however taken into account.

\begin{figure}
\vfill
\centerline{\mbox{\epsfysize=10.0truecm\epsffile{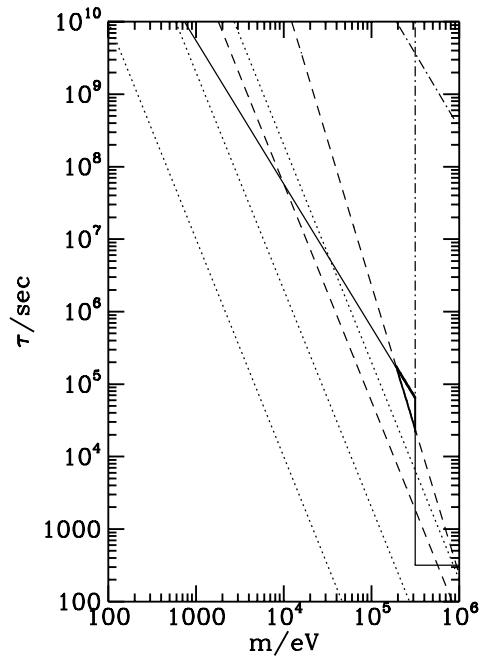}}}
\caption{\it Excluded region for cosmological reasons is delimited by
the continuous line;the dashed line corresponds to the possibility of
neglecting some statistical effect. The dotted lines corresponds to eq.
\protect\eqn{eq:mtv} for $V = 10^7, 10^8, 10^9$ GeV. }
\label{fig:parf}
\end{figure}

In most cases this can be done using eq. \eqn{eq:solser}, which
is a fair approximation when $f_{dy}(x,t)
\ll 1 - fer(x)$, for any $x \geq x_{dg}$. The condition
is most restrictive at the decoupling itself, where it reads
\be
{t_{dg} \over \tau} \ll 1.7 \times 10^{-2}
\left( m_{\nu_\tau} \over T_{dg} \right)^3
\ee
or, replacing the values to $T_{dg}$ and $t_{dg}$, known from standard
microphysics,
\be
{ \tau \over ~\mbox{\rm sec} }
\left( m_{\nu_\tau} \over 4 \times 10^3\, ~\mbox{\rm keV} \right)^3 \gg 1
\ee
In the plane $m_{\nu_\tau} - \tau$ the boundaries of the region where the
solution can be approximated by the \eqn{eq:solser}, are represented by the
lower dashed line (see fig.\ref{fig:parf}).
For parameter choices leading below this
curve no approximations can be used.
If, for any $x \geq x_{dg}$, the unequality $f_{dy}(x,t)+fer(x)\ll 1$ is
satisfied, we can use the exponential solution.
In this case, neutrino lifetime is long enough to allow a neglect of the
effects of fermi distribution. The unequality, written in terms of
$m_{\nu_\tau}$ and $\tau$, becomes:
\be
4.8 \times 10^{10} \zeta(3) \left( \tau \over \mbox{\rm sec}  \right)^{-1}
\left( m \over \mbox{\rm keV} \right)^{-3}
+{2 \over \exp\left[{2 \times
10^3 \left( m \over \mbox{\rm keV} \right)^{-1}}\right] +1 } \ll 1
\ee
This corresponds to the upper dashed--line in fig. \ref{fig:parf}.
Only in the area
above this line we can use the exponential solution, neglecting any kind
of statistical effect.
%%%%%%%%%%%%%%%%%FIGURE parameter plane {file= parf.ps}%%%%%%%%%%%%%%%%%
%%%%%%%%%%%%%%%%%%%%%%%%%%%%%%%%%%%%%%%%%%%%%%%%%

In fig. \ref{fig:parf}  we also show which parameter choices are excluded
for various
reasons. The area at the right and above the continuous line is excluded
in order not to modify the results of the BBNS
\cite{dgt:94} and in order not to cause drastic
changes to the power spectrum of density fluctuations
\cite{wgs:94}.
For permitted values of $m_{\nu_\tau}$ and $\tau$ we can make recourse to the
approximated solution (eq. \eqn{eq:solser}) in a few cases, while, in most
cases, the whole solution (eq. \eqn{eq:solfin}) is required.
We can also notice that, in the limit of cosmological permitted parameter,
the exponential solution can almost never be used.
The parameters which permit such a solution are only the ones in the marked
triangle in the figure.

According to familon models, once the scale $V$ of symmetry breaking is known,
mass and lifetime of $\nu_\tau$ decaying into a light neutrino $\nu_\mu$ and a
familon $\phi$ are related:
\bel{eq:mtv}
\tau \simeq 2 \times 10^6 \left( V \over 10^{10} ~\mbox{\rm GeV} \right)^2
\left( m_{\nu_\tau} \over \mbox{\rm keV}  \right)^{-3}  ~{\mbox{\rm yr}} ~.
\ee
Eq. \eqn{eq:mtv} yields the curves drawn in fig. \ref{fig:parf}
(dotted line) for $V = 10^7, 10^8, 10^9$ GeV.
These values of $V$ are consistent with current theoretical expectations. It is
also fair to outline that a self consistent picture arises from their choice.
The family symmetry, where spontaneous breaking generates the $\phi$ scalars,
is also consistent with the see--saw argument. Within this scheme, the lifetime
of $\tau$ neutrinos allows them to decay after BBNS and early enough to produce
a number of desirable cosmological features. Among them let us outline the
encreased energy of the low--mass (or massless) backgrounds. This causes a
later equivalence redshift (when the energy density of non--relativistic
components gets above that of relativistic ones) and therefore rises the
expected CMBR fluctuations in respect to fluctuations on the Mpc's scale. In
turn, this allows a value of the bias parameter $b>1$. An enhancement of
neutrino backgrounds has therefore several appealing features.

In this note we
have shown that, if this occurs, the expected neutrino energy distribution can
have non trivial features and we have provided the solutions of the equations
giving their details. The mechanism acts to delay the
decay of heavy neutrinos. Their yields will therefore suffer a smaller redshift
and a further enhancement of the decay background shall follow.

\section*{Acknowledgments}
We are pleased to thank Antonio Masiero, Marco Roncadelli and Daniela
Zanon  for helpful discussions.

\begin{Thebibliography}{99}

\bibitem{seb}
S. Ghigna, S. Borgani, S.A. Bonometto, L. Guzzo, A. Klypin, J. Primack,
R. Giovanelli, M.P. Haynes, \apj{437}{94}{L71}

\bibitem{bv:85}
S.A. Bonometto, R.Valdarnini, \apj{299}{85}{L71}

\bibitem{vb:85}
R.Valdarnini, S.A. Bonometto, \aa{146}{85}{235}

\bibitem{aos:85}
S. Achilli, F. Occhionero, R. Scaramella \apj{299}{85}{577}

\bibitem{hol:89}
J. Holtzmann \apjs{71}{89}{1}

\bibitem{pierpa}
E. Pierpaoli, S.A. Bonometto {\it Astron. \& Astroph}, (1994) in press

\bibitem{bgm:94}
S.A. Bonometto, F. Gabbiani, A. Masiero \pr{D49}{94}{3918}

\bibitem{phkc}
J.R. Primack, J.Holtzman, A. Klypin, D.O. Caldwell \nat{}{94}{}

\bibitem{gr:94}
 G.Gelmini, E. Roulet, preprint UCLA/94/TEP/36-- hep-ph/9412278

\bibitem{yana:79}
 T.Yanagida {\it proc.of the Workshop on Unified Theories and Baryon Number in
 the Universe. KEK. Japan} (1979)

\bibitem{gm:79}
 M.Gell--Mann, Ramond, R. Slansky {\it Supergravity} (North Holland 1979)

\bibitem{jrs:86}
 R. Johnson, S. Ranfone, J. Schecter, \pl{B179}{86}{355};\\
 R. Johnson, S. Ranfone, J. Schecter, \pr{B35}{87}{282}

\bibitem{glas:91}
 S.L. Glashow, \pl{B256}{91}{218}

\bibitem{GNY:83}
G.B.Gelmini, S.Nussinov, T.Yanagida \np{B219}{83}{31}

\bibitem{wil:82}
F.Wilczek, \prl{49}{82}{1549}

\bibitem{tao:94}
Z. Tao preprint hep-ph 9412223

\bibitem{wgs:94}
 M. White, G. Gelmini, J. Silk preprint astro-ph 9411098 CfPA-94-TH-48,
 UCLA/94/TEP/37

\bibitem{cmp}
 Y. Chikashige, R.N. Mohapatra, R.D. Peccei \pl{B98}{81}{265};\\
 Y. Chikashige, R.N. Mohapatra, R.D. Peccei \prl{45}{80}{1926}

\bibitem{schv:82}
 J. Schechter, J.W.F. Valle, \pr{D25}{82}{283E};\\
 J. Schechter, J.W.F. Valle, \pr{D25}{82}{774}

\bibitem{comp}
 G. Gelmini, J.W.F. Valle, \prl{B142}{84}{181};\\
 K.S.Babu, R.N.Mohapatra, \prl{66}{91}{556}

\bibitem{dgt:94}
 S.Dodelson, G.Gyuk, M.S.Turner {\it preprint} astro-ph/9312062,
 (submitted to {\it Physical Review D})

\end{Thebibliography}

\end{document}